\documentclass[aps,pra,preprint,showpacs]{revtex4}
\usepackage{bm}
\begin{document}
\title{Guage-field model of superfluid turbulence in the zero-temperature limit}

\author{Mohammad Mehrafarin}
\affiliation{Physics Department, Amirkabir University of Technology, Tehran 15914, Iran}
\email{mehrafar@aut.ac.ir}

\begin{abstract}
We present a gauge-field extension of the Bose condensate model that describes $T\approx0$ superfluid turbulence generated by the macroscopic motion of the superfluid.  We first establish that the condensate model is dual to the short-range interacting loop gas model, wherein the loops represent quantum vortex lines. Vortex lines form, interact and proliferate as a result of the superfluid motion. Our extension is based on incorporating the Biot-Savart interaction between vortex lines, which is lacking in the loop gas model. We show that the extended loop gas is dual to a Ginzburg-Landau model, wherein the gauge coupling is between the macroscopic velocity field of the superfluid and the condensate. Applying the model to cylindrical and pipe flows, we describe how turbulence transitions with and without intermediate vortex flow, respectively. 

\end{abstract}


\maketitle

\section{Introduction}

Superfluid turbulence can be generated by a number of methods, notably, by counterflow, or by macroscopic motion of the superfluid \cite{Kozik}. Contrary to the counterflow method which involves the normal component, externally imposed superfluid motion pertains to the zero temperature limit (nowadays often taken to mean $T\leq0.6 K$) we are considering. Vortices form in the shape of atomically thin vortex lines of quantized strength, a tangle of which creates the superfluid turbulent state. At $T\approx 0$, the fundamental microscopic theory of superfluids, based on many-body quantum mechanics, is difficult to use to study flow phenomena. The effective model of a weakly interacting Bose condensate provides a simpler approach \cite{Roberts}. It captures some essential physics of superfluid flow, in particular, the structure of the vortex lines. However, it does not explain the observed behavior, described below, for transition to turbulence caused by superfluid motion.

Superfluid transitions from laminar to turbulent flow with or without intermediate mixed (vortex) flows, depending on the geometry of the superfluid's macroscopic motion. Most notable examples are the cylindrical vessel (or Couette) flow and pipe flow, respectively \cite{Kagan,Barenghi}. In the superfluid cylindrical flow, as long as the angular velocity of the rotating vessel is below a first critical value $\Omega_1$, the fluid remains stationary. The intermediate mixed flow forms when the angular velocity is increased above $\Omega_1$, with the superfluid breaking into an ordered array of vortex lines that are aligned along the rotation axis. Above a second critical angular velocity $\Omega_2$ ($>\Omega_1$), the ordered vortex flow transitions into a turbulent flow in which quantum vortices invade the whole fluid by forming a chaotic tangle. In pipe flow, however, turbulent flow sets in when the superfluid velocity exceeds a critical value and there is no intermediate mixed flow. 

The correspondence of superfluid turbulence with superconductivity is intriguing \cite{Kagan}. External driving weakens the ability of a flow to remain laminar, just as external currents weaken the ability of a metal/alloy to be superconductive. In both examples above, expulsion of the macroscopic velocity from the liquid interior corresponds to the Meissner effect, which occurs below a critical value of the applied magnetic field. The ensuing flow can be divided into two types according to how the expulsion breaks down, corresponding to types I superconductivity and II. The array of vortices in the mixed cylindrical flow corresponds to the Abrikosov lattice of magnetic flux lines. The latter forms the mixed state of type II superconductors when the magnetic field lies between two critical values. If the field increases even more, the flux lines will invade the whole sample and superconductivity will be lost. In type I superconductivity, which corresponds to the pipe flow, this happens when the external field exceeds a critical value and there is no intermediate mixed state.

Vortex lines are closed, they form and proliferate as a result of the superfluid motion, tangling to destroy superflow.  Therefore, superfluid turbulent transition should be describable by a grand canonical ensemble of fluctuating loops of arbitrary size and shape (the loop gas) representing the atomically thin vortex lines. Corroboratively,
we first establish that the condensate model is dual to the short-range interacting loop gas model.  However, once created, vortex lines interact via the long range Biot-Savart (BS) potential that results from their self-induced velocity field \cite{Saffman}.  This interaction, which is a by-result of the macroscopic superfluid motion, is lacking in the loop gas,
thus rendering the condensate model improper for describing flows that potentially involve vortices. By incorporating the BS interaction, we show that the extended loop gas is dual to a Ginzburg-Landau (GL) model wherein the gauge coupling is between the macroscopic velocity filed of the superfluid and the condensate. 
The model coincides with the GL model of superconductivity \cite{de Gennes}, thus, explaining the peculiar similarity between the latter and superfluid turbulence. Applying the model to cylindrical and pipe flows, we describe how turbulence transitions with and without intermediate vortex flow, respectively.

\section{The condensate model and duality to the loop-gas model}

In the standard condensate model, the large number of condensate bosons per unit volume, $n_0$, gives rise to a collective irrotational velocity field, $\textbf{v}_0=\nabla \phi$, for the condensate, which corresponds to the superfluid velocity. The macroscopic wave function of the condensate is given by $\psi=\sqrt{n_0}\, {\text {exp}}(im \phi/\hbar)$, where $m$ is the boson mass. For a condensate of bosons with self-interaction of strength $g>0$, the equilibrium energy is given by the Gross-Pitaevskii energy functional \cite{Kleinert}
\begin{equation}
E_0[\psi,\psi^\star]=\int d^3x\,[\frac{\hbar^2}{2m}|\nabla \psi|^2- \mu |\psi|^2+\frac{g}{2} |\psi|^4] \label{1}
\end{equation}
where $\mu>0$ is the chemical potential near $T\approx 0$. The Gross-Pitaevskii (GP) equation that follows from the minimization of this functional possesses vortex line solutions (see e.g. \cite{Kleinert}); the core being an atomically thin hole surrounded by a macroscopic region of irrotational flow. Thus, the superfluid can lower its total energy by creating inhomogeneity in the flow, concentrating vorticity along highly localized regions.

Vortex lines form closed loops, they cannot start nor end in the fluid interior because the vorticity field is solenoidal. They form and proliferate as a result of the superfluid motion, tangling to destroy irrotational superflow.  Therefore, the superfluid turbulent transition should be describable by a grand canonical ensemble of fluctuating loops of arbitrary size and shape (the loop gas) representing the atomically thin vortex lines. (The loop gas has been applied by K. Kleinert \cite{Kleinert} to model both superfluid and superconductive phase transitions, where in the latter, loops represent magnetic flux lines. In contrast to ours, in Kleinert's model proliferation of loops is caused by temperature change near the transition temperature, which leads to the loss of order and signals phase transition.) Indeed, the condensate model (\ref{1}) is dual to the short-range interacting loop gas model.  This was originally shown by Kleinert \cite{Kleinert}, and our derivation, being tailor made for the problem at hand, is along similar lines (see Appendix A). The duality is established by mapping the partition function of the condensate model, namely,
\begin{equation}
W_0=\int d[\psi] d[\psi^\star]\, e^{-E_0/k_BT} \label{PF}
\end{equation}
into the following form ($\lambda=h/\sqrt{2\pi mk_BT}$ being the de Broglie thermal wavelength)
\begin{eqnarray}\begin{array}{c}
W_0=\sum_{N=1}^N \frac{1}{N!}\prod_{l=1}^N 
\{\int_0^\infty \frac{ds_l}{s_l}\,e^{\mu s_l/k_BT}\oint d[\bm{x}(s_l^\prime)] \exp \left(-\pi\lambda^{-2}\int_0^{s_l}ds_l^\prime\, \dot{\bm{x}}^2(s_l^\prime)\right)\}\\ \times \exp \left(-\frac{1}{k_BT}\sum_{l,k=1}^N U_0[\bm{x}(s_l^\prime),\bm{x}(s_k^{\prime\prime})]\right),\\ U_0[\bm{x}(s_l^\prime),\bm{x}(s_k^{\prime\prime})]= \frac{1}{2}g\int_0^{s_l}\int_0^{s_k}ds_l^{\prime} ds_k^{\prime\prime}\, \delta^3(\bm{x}(s_l^\prime)-\bm{x}(s_k^{\prime\prime}))\label{W0}
\end{array}
\end{eqnarray}
which represents the grand canonical partition function of a system of loops parametrized by the dimensionless variable $s_l^\prime$. The factor $1/s_l$ in the first integral reflects the fact that every value of $s_l^\prime$ in the interval $(0,s_l)$ can represent the start/end point of the loop, giving rise to $s_l$ indistinguishable configurations. The exponential following this factor is a Boltzmann factor favoring configurations with large $s_l$ values, i.e., large loop sizes, as occur in turbulent transition. The next exponential corresponds to the kinetic energy of the loops, and the last one represents the Boltzmann factor associated with the short range (delta-function) repulsive interaction between the loops. 

In the zero-temperature limit, only the minimum energy configuration counts in (\ref{PF}) so that the GP equation can be considered exact. Vortex lines, once created, interact via the long range Biot-Savart (BS) potential that results from their self-induced velocity field.  This interaction, which is a by-result of the superfluid motion, is lacking in the loop gas. In the next section, we extend the loop gas model by incorporating the BS interaction. 

\section{The extended loop gas model and duality to the GL model}
Vortex lines evolve and move under the influence of their self-induced solenoidal velocity field as well as the background irrotational superflow. The BS interaction between vortices is just the hydrodynamic energy of the self-induced velocity field. For loops parametrized by $s_l^\prime$ and $s_k^{\prime\prime}$ and carrying one quantum of circulation $\kappa=h/m$ each, the BS potential energy is given by
$$
U_{BS}[\bm{x}(s_l^\prime),\bm{x}(s_k^{\prime\prime})]=\int_0^{s_l}\int_0^{s_k}\frac{1}{2}\rho \kappa^2 \frac{\dot{\bm{x}}(s_l^\prime)\cdot \dot{\bm{x}}(s_k^{\prime\prime})}{4\pi|(\bm{x}(s_l^\prime)-\bm{x}(s_k^{\prime\prime})|}ds_l^\prime \,ds_k^{\prime\prime}
$$
where the parameter $\rho$ (with dimensions of mass density), to be fixed later, appears on dimensional grounds. The partition function of the extended loop gas, $W$, is obtained by including the Boltzmann factor associated with the BS interaction between all pairs of loops in (\ref{W0}). Thus
\begin{eqnarray}\begin{array}{c}
W=\sum_{N=1}^N \frac{1}{N!}\prod_{l=1}^N 
\{\int_0^\infty \frac{ds_l}{s_l}\,e^{\mu s_l/k_BT}\oint d[\bm{x}(s_l^\prime)] \exp \left(- \pi\lambda^{-2}\int_0^{s_l}ds_l^\prime\,\dot{\bm{x}}^2(s_l^\prime)\right)\}\\ \qquad\times \exp \left(-\frac{1}{k_BT}\sum_{l,k=1}^N (U_0[\bm{x}(s_l^\prime),\bm{x}(s_k^{\prime\prime})]
+ U_{BS}[\bm{x}(s_l^\prime),\bm{x}(s_k^{\prime\prime})])\right )\label{BW}.
\end{array}
\end{eqnarray}
Using the auxiliary field $\bm{A}(\bm{x})$, this Boltzmann factor can be cast into the form (see Appendix B)
\begin{equation}
\int d[\bm{A}]\Phi[\bm{A}]\exp \left(- \frac{\rho}{k_BT}\left\{\int d^3x\, \frac{1}{2}(\nabla\times \bm{A})^2+i\kappa\sum_{l=1}^N\int_0^{s_l}ds_l^\prime \,\dot{\bm{x}}(s_l^\prime) \cdot \bm{A}(\bm{x}(s_l^\prime))\right\}\right).\label{gauge}
\end{equation}
(The functional $\Phi[\bm{A}]$ is a gauge fixing factor introduced in view of the gauge invariance of the integrand. $\Phi$ restricts the integral to the physical values of the field $\bm{A}$ and thus renders them finite. It can be chosen from a variety of gauge fixing factors corresponding to different choices of the gauge, but the result will be independent of that choice, of course.) Therefore, (\ref{BW}) becomes
\begin{eqnarray}\begin{array}{l}
W=\int d[\bm{A}]\Phi[\bm{A}]\exp \left( -\frac{\rho}{k_BT} \int d^3x\, \frac{1}{2}(\nabla\times \bm{A})^2\right ) \\ \qquad
\times \sum_{N=1}^N \frac{1}{N!}\prod_{l=1}^N 
\{\int_0^\infty \frac{ds_l}{s_l}\,e^{\mu s_l/k_BT}\oint d[\bm{x}(s_l^\prime)] \exp \left(- \pi\lambda^{-2}\int_0^{s_l}ds_l^\prime\,\dot{\bm{x}}^2(s_l^\prime)\right)\}\\ \qquad\qquad
\times \exp\left(-\frac{1}{k_BT}\sum_{l,k=1}^NU_0[\bm{x}(s_l^\prime),\bm{x}(s_k^{\prime\prime})] -i\frac{\rho\lambda^2}{\hbar}\sum_{l=1}^N\int_0^{s_l}ds_l^\prime\, \dot{\bm{x}}(s_l^\prime) \cdot \bm{A}(\bm{x}(s_l^\prime)) \right) \label{BC}
\end{array}
\end{eqnarray}
Hence (see Appendix C) 
\begin{eqnarray}\begin{array}{l} 
W=\int d[\bm{A}]\Phi[\bm{A}]\exp \left( -\frac{\rho}{k_BT} \int d^3x\, \frac{1}{2}(\nabla\times \bm{A})^2\right ) \\ \qquad
\times\int d[\psi] d[\psi^\star]\exp \left(-\frac{1}{k_BT} \int d^3x\, [\frac{\hbar^2}{2m}|(\nabla-i\frac{\rho\lambda^2}{\hbar}\bm{A}) \psi|^2- \mu |\psi|^2+\frac{g}{2} |\psi|^4] \right)\label{C}
\end{array}
\end{eqnarray}
where $\bm{A}$ embodies the vortex BS interactions.
Let us introduce $\bm{V}=\bm{A}/\lambda$, which has the dimensions of velocity, and let  $\rho\lambda^3=m$. Because BS interactions are  a by-result of the superfluid motion, we identify $\bm{V}$ as the macroscopic velocity field of the superfluid, in the absence of which the condensate model is restored. Then
\begin{eqnarray}\begin{array}{c}
W=\int d[\psi] d[\psi^\star]d[\bm{V}]\Phi[\bm{V}]\, e^{-E/k_BT},\\
E[\psi,\psi^\star,\textbf{V}]
=\int d^3x\, [\frac{\hbar^2}{2m}|(\nabla-i\frac{m}{\hbar}\bm{V}) \psi|^2- \mu |\psi|^2+\frac{g}{2} |\psi|^4 +\frac{1}{2}\frac{m}{\lambda}\, (\nabla \times \bm{V})^2 ] \label{GL}
\end{array}
\end{eqnarray}
viz, the extended loop gas model is dual to the GL model. (We should mention that a similar GL functional has been derived long ago \cite{Chelaflores} by merely demanding local gauge invariance for the Gross-Pitaevskii energy functional, wherein $\bm{V}$ is attributed to the depletion velocity and no connection is made with turbulence.)

The GL functional (\ref{GL}) is invariant under local gauge transformations $\bm{V}\rightarrow\bm{V}+\nabla\Lambda,\, \psi\rightarrow\psi e^{im\Lambda/\hbar}$. The spontaneous breaking of the local guage symmetry leads to the Higgs mode corresponding to the superfluid Meissner effect:
The Goldstone mode associated with the spontaneous breaking of the global gauge symmetry of (\ref{1}) corresponds to the long-wavelength perturbations of the velocity potential. The condensate flows irrotationally with velocity given by the gradient of the perturbation field. When set into motion, because of the gauge coupling of the macroscopic velocity to the condensate, the long range order becomes costly and the Goldstone mode disappears. Because $\mu>0$, it transforms to a short range (Higgs) mode, giving way to a finite penetration depth for the external velocity, which corresponds to the Meissner effect. This signals the onset of turbulence through vortex production.

The field equations that follow from the minimization of functional $E$ are the well-known GL equations
\begin {eqnarray}\begin{array}{c}
-\frac{\hbar^2}{2m}(\nabla-i\frac{m}{\hbar}\bm{V})^2 \psi+ g|\psi|^2\psi=\mu\psi    \\
\frac{m}{\lambda} \nabla \times\nabla \times\bm{V}=\frac{i\hbar}{2} (\psi\nabla\psi^\star-\psi^\star\nabla\psi)-m\bm{V}|\psi|^2.\label{GLE} 
\end{array}
\end{eqnarray}
These equations, which replace the GP equation, describe turbulent transition caused by the macroscopic motion of the superfluid. 

\section{Transition to turbulence}
The GL model of superfluid turbulence (\ref{GL}) explains the similarity, discussed in the Introduction, between this phenomena and superconductivity. The expulsion of the external macroscopic velocity, which occurs below a critical driving value, is just the superfluid Meissner effect. The Meissner effect breaks down when the macroscopic velocity is too large. Superfluid flow can be divided into two classes according to how this breakdown occurs. In type I flows (the pipe flow), laminar superflow is abruptly destroyed when the velocity rises above a critical value. In type II flows (the cylindrical flow), raising the velocity above a first critical value leads to an intermediate mixed flow, in which an increasing amount of vorticity penetrates the superfluid through ordered arrays of quantized vortices. At a second critical value, superflow is completely destroyed.

To see how the above scenario works for turbulent transitions, we apply the GL equations (\ref{GLE}) to a superfluid subject to external motion. Starting from turbulent/vortex flow, if we continuously decrease the velocity, at a certain critical value superflow begins to form. Two situations arise according to the two distinct solutions of the GL equations:

(i) A condensate phase appears abruptly in the bulk of the fluid, where $|\psi|$ is the same at all points. Such a condensate corresponds to laminar superflow, which entails decoupling from the macroscopic motion when the velocity is sufficiently reduced. Working in the gauge where $\psi$ is real, the  GL equations read:
\begin {eqnarray}\begin{array}{c}
\frac{1}{2}mV^2+g\psi^2=\mu  \\
\nabla^2\bm{V}=\lambda\bm{V}\psi^2.
\end{array} 
\end{eqnarray}
The first equation gives $V$ $< V_1=\sqrt{2\mu/m}$, which expresses low driving kinetic energy. Its first order solution $\psi=\sqrt{\mu/g}$ yields the London-like equation $\xi ^2\nabla^2\bm{V}=\bm{V}$,
where $\xi=\sqrt{g/\mu \lambda}$ is the penetration length for the external velocity, beyond which it decouples from the superflow. The London equation explains the Meissner effect, which takes place at the critical velocity $V_1$ as the flow enters the condensate phase of laminar superflow. In terms of the vortex core radius $a_0=\hbar/\sqrt{2m\mu}\sim 3\text{\AA}$, we have $V_1=\hbar/ma_0\sim 54$ m/s, comparable to the Landau critical velocity ($\approx 58$ m/s).

(ii) A condensate phase appears in the bulk of the fluid by spontaneous nucleation of condensate regions. In regions where nucleation occurs, superflow is just beginning to appear and, therefore, $|\psi|$ is small. Since $|\psi|$ is not the same at all points, such a condensate is distinct from laminar superflow (i); it corresponds to mixed vortex flow. The GL equations can be linearized to give the Shr\"{o}dinger-like equation
\begin{equation}
-\frac{\hbar^2}{2m}(\nabla-i\frac{m}{\hbar}\bm{V})^2 \psi=\mu\psi. \label{shrod}
\end{equation}
Let us ignore boundary and consider an infinite domain.

For pipe flow, where $\bm{V}$ is along the pipe axis, vortex flow is not possible because the solutions of (\ref{shrod}) satisfy $|\psi|=\text{const}$. In other words, turbulent pipe flow transitions into laminar superflow without the intermediate vortex flow. Meissner effect takes place at the critical velocity $V_1$ and the flow is of type I. 

For cylindrical flow, where $\bm{V}=\bm{\Omega} \times{\bm{r}}$, a condensate phase corresponding to an eigensolution of (\ref{shrod}) has eigenvalue $\mu=(2n+1)\hbar \Omega \label{ev}$, 
where $n$ is a nonnegative integer. Vortex flow, thus, appears for angular velocities $\Omega<\Omega_2=\mu/\hbar=\hbar/2ma_0^2\sim 10^{11}$ rad/s. (The upper bound $\Omega_2$ is practically unachievable. Its significance is theoretical so as to establish a type II flow.) By still decreasing the angular velocity, the number of vortices decreases until at some critical value $\Omega_1$ there remains only one vortex line at the center of the cylinder ($r=0$). The superflow of this vortex line has azimuthal velocity $\bm{v}= \hat{\bm{\varphi}}\hbar/mr$, and the superfluid density is the same at all points except along the vortex line. $\Omega_1$ can be determined from $E=L\Omega_1$, where $E$ is the kinetic energy of the superflow and $L$ is its angular momentum. Hence
$$
\Omega_1=\frac{1}{2} \frac{\int  v^2 d^3x}{\int rv\,d^3x}=\frac{\hbar}{m(R^2-a_0^2)} \ln\frac{R}{a_0}\approx \frac{\hbar}{mR^2} \ln\frac{R}{a_0}
$$
where $R$ is the radius of the cylinder. For a typical vessel with  $R\sim 1$ mm, we have $\Omega_1\sim 0.2$ rad/s. Below $\Omega_1$, there is no vortex line and the condensate phase of laminar superflow forms. Therefore, the flow is of type II. 

Superfluid vortex dynamics has been studied via numerical simulations of the GP and projected GP equations \cite{Berloff}. It would be instructive to undertake similar numerical studies based on the LG equations (\ref{GLE}). Comparison of the results would also test of our model.

\section{Recapitulation}

We have presented a guage-field extension of the Bose condensate model that describes $T\approx 0$ superfluid turbulence generated by the externally imposed macroscopic motion of the superfluid.  As a result of the superfluid motion, vortex loops form and proliferate, leading to the  loss of ordered flow. Thus, the superfluid turbulent transition should be describable by the loop gas model, where the loops represent the quantum vortex lines. Corroboratively, we first established that the condensate model is dual to the short-range interacting loop gas model. However, once created, vortex lines interact via the BS potential, a by-result of the superfluid motion that is lacking in the loop gas model. By incorporating the BS interaction, we have shown that the extended loop gas model is dual to a GL model wherein the gauge coupling is between the macroscopic velocity filed of the superfluid and the condensate. The model coincides with the GL model of superconductivity, thus, explaining the similarity between the latter and superfluid turbulence. Applying the model to cylindrical and pipe flows, we have described how turbulence transitions with and without intermediate vortex flow, respectively. 

\appendix 
\section{}
Let us first derive the following result, which we shall be needing in the sequel:
\begin{equation}
\det(-\frac{\hbar^2}{2m}\nabla^2+U)^{-1}=\exp (\int_0^\infty \frac{dt}{t}\oint  d[\bm{x}(t^\prime)]\exp \left(-\frac{1}{\hbar}\int_0^{t}dt^\prime [\frac{1}{2}m\dot{\bm{x}}^2(t^\prime)+U(\bm{x}(t^\prime))] \right)). \label{A1}
\end{equation}
The Green function of the imaginary time ($it\rightarrow t$) Shr\"{o}dinger equation satisfies
$$
[\hbar\partial_t-\frac{\hbar^2}{2m}\nabla^2+U(\bm{x})]G(\bm{x},t)=\hbar\delta^3(\bm{x})\delta(t).
$$
Using the integrating factor $I(t)=\exp[\frac{t}{\hbar}(-\frac{\hbar^2}{2m}\nabla^2+U)]$, the above equation gives $IG=\delta^3(\bm{x})\int_0^tdt^\prime\delta(t^\prime)I(t^\prime)=\delta^3(\bm{x})$, or
\begin{eqnarray}\begin{array}{cl}
G(\bm{x},t)&=\exp[-\frac{t}{\hbar}(-\frac{\hbar^2}{2m}\nabla^2+U)] \delta^3(\bm{x})\\
&=\int_{\bm{x(0)}=0}^{\bm{x}(t)=\bm{x}} d[\bm{x}(t^\prime)]\exp \left(-\frac{1}{\hbar}\int_0^{t}dt^\prime[ \frac{1}{2}m\dot{\bm{x}}^2(t^\prime)+ U(\bm{x}(t^\prime))] \right).\label{A2}
\end{array}
\end{eqnarray}
The last line expresses the standard path integral formulation. Using $\int_\epsilon^\infty \frac{du}{u}e^{-u}=\ln\epsilon^{-1}$, where
$\epsilon\rightarrow 0^+$, we have
$$
\int_0^\infty \frac{dt}{t}\,G(\bm{x},t)=\ln(-\frac{\hbar^2}{2m}\nabla^2+U)^{-1}\delta^3(\bm{x})
$$
to within an additive (divergent) constant. Hence
\begin{equation}
\text{Tr}\ln(-\frac{\hbar^2}{2m}\nabla^2+U)^{-1}=\int_0^\infty \frac{dt}{t}\,\text{Tr}\,G(\bm{x},t)=\int_0^\infty \frac{dt}{t}\,G(0,t) \label{A3}
\end{equation}
as
$$
\text{Tr}\,G(\bm{x},t)=\int \frac{d^3k}{(2\pi)^3} \tilde{G} (\bm{k},t)=G(0,t)
$$
$\tilde{G}$ being the Fourier transform of $G$. Thus, (\ref{A1}) follows from (\ref{A3}) on using $\text{Tr}\ln B=\ln \det B$ and (\ref{A2}).

Now define
$$
E_0=\int d^3x\, (\frac{\hbar^2}{2m}|\nabla \psi|^2+ U |\psi|^2+\frac{g}{2}|\psi|^4).
$$
It follows from
$$
\int d[\phi] \exp \left(-\frac{1}{k_BT}\int d^3x\,(\sqrt{\frac{2}{g}}\,\phi-i\sqrt{\frac{g}{2}}\,|\psi|^2)^2\right)=1
$$
that
$$
\exp \left(-\frac{g}{2k_BT} \int d^3x\,|\psi|^4 \right)=\int d[\phi] \exp\left(-\frac{2}{k_BT} \int d^3x\,(\frac{1}{g}\,\phi^2-i\phi|\psi|^2)\right)
$$
and so
\begin{eqnarray}\begin{array}{c}
\int d[\psi] d[\psi^\star]\, e^{-E_0/k_BT}=\int d[\psi] d[\psi^\star] d[\phi]\,e^{-H/k_BT}\exp\left(-\frac{2}{k_BT} \int d^3x\,\frac{1}{g}\,\phi^2\right),\\
H=\int d^3x\, (\frac{\hbar^2}{2m}|\nabla\psi|^2+U|\psi|^2-2i\phi|\psi|^2)=\int d^3x\, \psi^\star(-\frac{\hbar^2}{2m}\nabla^2+U-2i\phi) \psi.\label{AA}
\end{array}
\end{eqnarray}
But
$$
\int d[\psi] d[\psi^\star]\, e^{-H/k_BT}=\det[-\frac{\hbar^2}{2m}\nabla^2+U-2i\phi]^{-1}=e^J
$$
where, from (\ref{A1}),
$$
J=\int_0^\infty \frac{dt}{t}\oint d[\bm{x}(t^\prime)]\exp \left(-\frac{1}{\hbar}\int_0^{t}dt^\prime [\frac{1}{2}m\dot{\bm{x}}^2(t^\prime)+ U(\bm{x}(t^\prime))-2i\phi(\bm{x}(t^\prime))] \right).
$$
Thus, (\ref{AA}) becomes,
\begin{eqnarray}\begin{array}{l}
\int d[\psi] d[\psi^\star]\, e^{-E_0/k_BT}=\int d[\phi]\, e^J\exp\left(-\frac{2}{k_BT} \int d^3x\,\frac{1}{g}\,\phi^2\right)=\\ \sum_{N=1}^N \frac{1}{N!}\prod_{l=1}^N 
\{\int_0^\infty \frac{dt_l}{t_l}\oint d[\bm{x}(t_l^\prime)]\exp \left(-\frac{1}{\hbar}\int_0^{t_l}dt_l^\prime\,[ \frac{1}{2}m\dot{\bm{x}}^2(t_l^\prime)+U(\bm{x}(t_l^\prime)) ] \right)\}\\ \qquad\times \int d[\phi]\exp\left(- \frac{2}{k_BT}\int d^3x\,\frac{1}{g}\phi^2+\frac{2i}{\hbar}\sum_{l=1}^N\int_0^{t_l}dt_l^\prime\, \phi(\bm{x}(t_l^\prime))\right)\label{A6}
\end{array}
\end{eqnarray}
having expanded the exponential $e^J$. Denoting the $\phi$ integral in the last line above by $F$, we can write
$$
F=\int d[\phi] \exp\left(-\int d^3x\, [\frac{2}{k_BT}\frac{1}{g}\phi^2-\frac{2i}{\hbar}\phi \sum_{l=1}^N\int_0^{t_l}dt_l^\prime\, \delta^3(\bm{x}-\bm{x}(t_l^\prime))]\right).
$$
It follows from
$$
\int d[\phi] \exp\left(-\int d^3x\,[\sqrt{\frac{2}{gk_BT}}\phi-\frac{i}{\hbar}\sqrt{\frac{gk_BT}{2}}\sum_{l=1}^N\int_0^{t_l}dt_l^\prime\, \delta^3(\bm{x}-\bm{x}(t_l^\prime))]^2\right)=1
$$
that
$$
F=\exp\left(-\frac{1}{\hbar^2}\frac{gk_BT}{2}\sum_{l,k=1}^N\int_0^{t_l}\int_0^{t_k}dt_l^\prime dt_k^{\prime\prime}\, \delta^3(\bm{x}(t_l^\prime)-\bm{x}(t_k^{\prime\prime}))]\right).
$$
Hence, (\ref{A6}) reads:
\begin{eqnarray}\begin{array}{l}
\int d[\psi] d[\psi^\star]\, e^{-E_0/k_BT}=\\\sum_{N=1}^N \frac{1}{N!}\prod_{l=1}^N 
\{\int_0^\infty \frac{dt_l}{t_l}\oint d[\bm{x}(t_l^\prime)] \exp \left(-\frac{1}{\hbar}\int_0^{t_l}dt_l^\prime\,[ \frac{1}{2}m\dot{\bm{x}}^2(t_l^\prime)+ U(\bm{x}(t_l^\prime)]\right)\}\\ \qquad \times \exp\left(-\frac{1}{\hbar^2}\frac{gk_BT}{2}\sum_{l,k=1}^N\int_0^{t_l}\int_0^{t_k} dt_k^{\prime\prime}\, \delta^3(\bm{x}(t_l^\prime)-\bm{x}(t_k^{\prime\prime}))\right) .\label{final}
\end{array}
\end{eqnarray}
Finally, putting $U=-\mu$ and $t=\frac{\hbar}{k_BT}s$, we obtain expression (\ref{W0}) for the partition function.

\section{}
Let us denote expression (\ref{gauge}) by $I$. Working in the guage $\nabla \cdot \bm{A}=0$, we have $(\nabla \times \bm{A})^2=-\bm{A}\cdot \nabla^2 \bm{A}$,  and so 
\begin{eqnarray}\begin{array}{cl}
I&=\int d[\bm{A}]\exp \left(- \frac{\rho}{k_BT}\left\{\int d^3x\, \frac{1}{2}\bm{A}\cdot (-\nabla^2) \bm{A}+i\kappa\sum_{l=1}^N\int_0^{s_l}ds_l^\prime \,\dot{\bm{x}}(s_l^\prime) \cdot \bm{A}(\bm{x}(s_l^\prime))\right\}\right)\\ 
&=\int d[\bm{A}]\exp \left(- \frac{\rho}{k_BT}\int d^3x\, [\frac{1}{2}\bm{A}\cdot (-\nabla^2) \bm{A}+i\sum_{l=1}^N\bm{\omega}_l\cdot \bm{A}]\right)\nonumber
\end{array}
\end{eqnarray}
where $\bm{\omega}_l(\bm{x})$ is the vorticity along loop $l$. The expression inside the square bracket can be written as
$$
\frac{1}{2}(\bm{A}-i\sum_{l=1}^N\bm{\omega}_l\nabla^{-2})\cdot (-\nabla^2) (\bm{A}-i\nabla^{-2}\sum_{k=1}^N\bm{\omega}_k)-\frac{1}{2}\sum_{l,k=1}^N\bm{\omega}_l\cdot \nabla^{-2} \bm{\omega}_k
$$
so that  
\begin{eqnarray}\begin{array}{c}
I=[\det(-\nabla^2)]^{-3/2}\exp\left( \frac{\rho}{2k_BT}\sum_{l,k=1}^N\int \int d^3x\, d^3x^\prime\, \bm{\omega}_l(\bm{x}) \cdot \nabla^{-2}\delta^3(\bm{x}-\bm{x}^\prime)\bm{\omega}_k(\bm{x}^\prime) \right)\propto\\
\exp\left(-\frac{\rho}{2k_BT} \sum_{l,k=1}^N\int \int d^3x\, d^3x^\prime\,\frac{\bm{\omega}_l(\bm{x}) \cdot \bm{\omega}_k(\bm{x}^\prime)}{4\pi|\bm{x}-\bm{x}^\prime|} \right)=\exp\left(-\frac{\rho\kappa^2}{2k_BT} \sum_{l,k=1}^N\int_0^{s_l} \int_0^{s_k} ds_l^\prime\, ds_k^{\prime\prime}\,\frac{\dot{\bm{x}}(s_l^\prime)\cdot \dot{\bm{x}}(s_k^{\prime\prime})}{4\pi|(\bm{x}(s_l^\prime)-\bm{x}(s_k^{\prime\prime})|} \right).\nonumber
\end{array}
\end{eqnarray}
Hence 
$$
I=\exp(-\frac{1}{k_BT}\sum_{l,k=1}^NU_{BS}[\bm{x}(s_l^\prime),\bm{x}(s_k^{\prime\prime})])
$$
to within an unimportant normalization constant. 

\section{}
In presence of the gauge field $\bm{A}$ with coupling constant $q$,
$\nabla\rightarrow \nabla-\frac{i}{\hbar}q \bm{A}$, and $U\rightarrow U+q\dot{\bm{x}}\cdot\bm{A}$ in the Lagrangian. Thus, going to imaginary time, (\ref{A1}) becomes
\begin{eqnarray}\begin{array}{l}
\det[-\frac{\hbar^2}{2m}(\nabla-\frac{i}{\hbar}q \bm{A})^2+U]^{-1}=\\ \exp (\int_0^\infty \frac{dt}{t}\oint  d[\bm{x}(t^\prime)]\exp \left(-\frac{1}{\hbar}\int_0^{t}dt^\prime [\frac{1}{2}m\dot{\bm{x}}^2(t^\prime)+U(\bm{x}(t^\prime))+iq\dot{\bm{x}}(t^\prime)\cdot\bm{A}(\bm{x}(t^\prime))] \right)). \nonumber
\end{array}
\end{eqnarray}
Also $E_0\rightarrow E$, where
$$
E=\int d^3x\, [\frac{\hbar^2}{2m}|(\nabla-\frac{i}{\hbar}q \bm{A}) \psi|^2+ U |\psi|^2+\frac{g}{2}|\psi|^4].
$$
Following the steps of Appendix A, we just have to make the replacement $U\rightarrow U+iq\dot{\bm{x}}\cdot\bm{A}$ in the final result (\ref{final}) to get
\begin{eqnarray}\begin{array}{l}
\int d[\psi] d[\psi^\star]\, e^{-E/k_BT}=\\ \sum_{N=1}^N \frac{1}{N!}\prod_{l=1}^N 
\{\int_0^\infty \frac{dt_l}{t_l}\oint d[\bm{x}(t_l^\prime)] \exp \left(-\frac{1}{\hbar}\int_0^{t_l}dt_l^\prime\,[ \frac{1}{2}m\dot{\bm{x}}^2(t_l^\prime)+ U(\bm{x}(t_l^\prime)]\right)\}\\ \qquad \times \exp\left(-\frac{1}{\hbar^2}\frac{gk_BT}{2}\sum_{l,k=1}^N\int_0^{t_l}\int_0^{t_k} dt_k^{\prime\prime}\, \delta^3(\bm{x}(t_l^\prime)-\bm{x}(t_k^{\prime\prime}))-\frac{i}{\hbar}q\sum_{l=1}^N\int_0^{t_l}dt_l^\prime\,\dot{\bm{x}}(t_l^\prime)\cdot\bm{A}(\bm{x}(t_l^\prime))\right).\nonumber
\end{array}
\end{eqnarray}
Putting $U=-\mu, q=\rho\lambda^2,t=\frac{\hbar}{k_BT}s$, we have
\begin{eqnarray}\begin{array}{l}
\int d[\psi] d[\psi^\star]\, e^{-E/k_BT}=\\ \sum_{N=1}^N \frac{1}{N!}\prod_{l=1}^N 
\{\int_0^\infty \frac{ds_l}{s_l}\,e^{\mu s_l/k_BT}\oint d[\bm{x}(s_l^\prime)] \exp \left(- \pi\lambda^{-2}\int_0^{s_l}ds_l^\prime\,\dot{\bm{x}}^2(s_l^\prime)\right)\}\\ \qquad
\times \exp\left(-\frac{1}{k_BT}\sum_{l,k=1}^NU_0[\bm{x}(s_l^\prime),\bm{x}(s_k^{\prime\prime})] -i\frac{\rho\lambda^2}{\hbar}\sum_{l=1}^N\int_0^{s_l}ds_l^\prime\, \dot{\bm{x}}(s_l^\prime) \cdot \bm{A}(\bm{x}(s_l^\prime)) \right). \label{C1}
\end{array}
\end{eqnarray}
Equation (\ref{C}) follows from (\ref{BC}) on using (\ref{C1}).

\end{document}